\newcommand{\E}{\mbox{ e}}
\newcommand{\ptd}{\partial}
\newcommand{\lt}{\left}
\newcommand{\rt}{\right}
\newcommand{\no}{\nonumber}

\newcommand{\ket}[1]{| #1 \rangle}

\newcommand{\av}[1]{\langle #1 \rangle}
\newcommand{\be}{\begin{equation}}
\newcommand{\ee}{\end{equation}}
\newcommand{\bea}{\begin{eqnarray}}
\newcommand{\eea}{\end{eqnarray}}
\newcommand{\rf}[1]{(\ref{#1})}

 \documentstyle[aps,eqsecnum,pra]{revtex}

\title{WKB for a damped spin}
 
\author{Petr A.~Braun}
 \address{Department of Theoretical Physics,
  Institute of Physics,
  Saint-Petersburg University,
  Saint-Petersburg, 198904 Russia}

\author{Daniel Braun, Fritz Haake}
\address{Fachbereich Physik,
  Universit\"at-Gesamthochschule Essen,
  45117 Essen, Germany}

\begin{document}
\maketitle
\begin{abstract}
The master equation for a damped spin well known from the theory of
 superradiance, is written as a finite-difference equation and  
 solved by a WKB-like method. The propagator thus obtained looks like the
 van Vleck propagator of a certain classical Hamiltonian system with one
 degree of freedom. A new interpretation is provided of the temporal
 broadening of initially sharp probability distributions as the analogue of
 the spreading of the quantum mechanical wave packet.

PACS numbers: 42.50F, 03.65.Sq
\end{abstract}

\section{Introduction}

This is the second of a series of papers concerned with dissipative
motion of a large spin which may be realized with many identical
collectively radiating two-level atoms. The large spin (the ``Bloch vector''
of quantum optics) has conserved length such that its classical state can be
described by two angles $\theta$ and $\varphi$. The classical dynamics is
that of an overdamped pendulum,
\bea
\dot{\theta}&=&\sin\theta\,, \no\\
\tan\frac{\theta(\tau)}{2}&=&{\rm e}^\tau
\tan\frac{\theta(0)}{2}\,,
\label{pendulum}\\
\varphi(\tau)&=&{\rm const}\,.\no
\eea
In geodesic jargon we could think of $\theta$ and $\varphi$ as defining
latitude and longitude and speak of creeping motion towards the south pole
along a great circle.

The starting point of the first
paper \cite{bbhw}, which we shall refer to as I, was the exact solution
of the ``superradiance master equation'' (see below) in the form of a
Laplace integral 
\cite{Haake3}. We evaluated that integral in a saddle-point
approximation and derived uniform asymptotics of the dissipative
propagator.

In the present paper we employ a different strategy far less
dependent on the specific properties of the problem. We observe
that in the limit of a large number $N$ of atoms the master equation
becomes a finite-difference equation with a small step, amenable to
solution by an approximation of the WKB type. The propagator solution
thus obtained takes the form of a van Vleck propagator involving the
action of a certain classical Hamiltonian system with one degree of
freedom. We find the pertinent Hamilton equations to be equivalent to
the saddle-point equation of I. The WKB approximation entails a new
interpretation of the temporal broadening of initially sharp probability
distributions as the analogue of the spreading of a quantum mechanical
wave packet.

As in I we shall employ the basis formed by the eigenstates $|jm\rangle$
of ${\bf J}^2$ and $J_z$ with the respective eigenvalues $j(j+1)$ and
$m$. The quantum number $j$ can take on integer or half integer
positive values (up to half the number $N$ of two-level atoms) and for fixed
$j$ the quantum number $m$ runs in unit steps from $-j$ to $+j$.
Denoting the density matrix elements by $\langle
j,m+k|\rho(\tau)|j,m-k\rangle=\rho_m^k(\tau)$ we can write the master
equation under study as
\be
J\frac{d\rho_m^k}{d\tau}=\sqrt{g_{m+k+1}g_{m-k+1}}\rho_{m+1}^k-
(g_m-k^2)\rho_m^k\label{master}\,,
\ee
where $\tau$ is a suitably scaled dimensionless time and $g_m$ denotes
the ``rate function''
\be
g_m=j(j+1)-m(m-1) \,.
\label{rate}
\ee
The particular solution $\rho_m^k(\tau)$ satisfying the
initial condition $\rho^k_m(\tau=0)=\delta_{mn}$ with a certain $n$ is
called the dissipative propagator and denoted as $D^k_{mn}(\tau)$. The
solution for an arbitrary initial density matrix then is
$\rho_m^k(\tau)=\sum_{n=-j}^j D^k_{mn}(\tau)\rho_n^k(0)$.

As shown in I, the master equation does not couple density matrix elements with
different skewness $k$. In particular, the probabilities
($k=0$) can be solved for independently of the
coherences ($k\ne 0$). The elements of the
``coherence propagator'' $D^k_{mn}$ with $k\ne 0$ are connected by an
elementary relation with those of the density propagator $D^0_{mn}$ [see
below, \rf{identity}]. When dealing with the probabilities $\rho_m^0$ and
their propagator $D_{mn}^0$ we shall drop the superscript $k=0$.

\section{Semiclassical asymptotics}
  
\paragraph {Finite-difference equation ---}
We are interested in the limit  of a large  number of atoms or, more
specifically, of a large spin,
\be
\sqrt{j(j+1)}\approx j+1/2\equiv J\gg1
\label{}
\ee
In that limit it is convenient to introduce a new independent variable 
$\mu$ and its increment $\Delta$ as 
\be
\mu=m/J\mbox{, }  \Delta=J^{-1}\,.   
\label{defmu} 
\ee
In the classical limit $\mu$ would become continuous in the range
$-1\ldots 1$. For our semiclassical perspective $\mu$ remains discrete
but neighboring values are separated by $\Delta$.
The master equation  (\ref{master}) for the densities (case $k=0$)
becomes a finite difference equation, 
\be
{\partial \rho  \left( \mu  ,\tau \right)  \over \partial \tau}= J
 \left[  
 \left( 1-\mu^2-\mu \Delta -\frac {\Delta^2}4 \right) \rho  \left( \mu
 +\Delta ,\tau \right)  
- \left( 1-\mu^2+\mu\Delta -\frac {\Delta^2}4  \right) \rho  \left( \mu
,\tau \right)  \right]\,.
\label{pab4}\ee 

\paragraph{WKB ansatz ---}
The WKB formalism for finite-difference equations with a small step is
well established. The general theory has been worked out mostly by Maslov 
\cite{Maslov1} whose notations we shall adhere to. The
WKB method for ordinary second-order difference equations was extensively
used to study the eigenvalues of huge
tridiagonal matrices occuring in the theory of Rydberg atoms in
external fields \cite{Braun1}. Closer to our topic, the leading
(exponential) term in the semiclassical solution of  master
equations of the type (\ref{pab4}) was obtained in \cite{Kitahara1}. We
follow the same lines but go a step further by also establishing
the preexponent, as is indeed necessary to get meaningful results
for most quantities of interest.

Let us look for a solution of (\ref{pab4}) in a form reminiscent of
the WKB wave function in a classically
forbidden domain,
\begin{equation} 
\rho  \left( \mu ,\tau \right) =A \left( \mu ,\tau \right) \E^{JS \left( 
\mu ,\tau \right) }\,.
\label{pab5}\end{equation} 
Here the prefactor $A$ and the ``action'' $S$ are smooth functions
satisfying the initial conditions  
\begin{equation}
S \left( \mu ,0 \right) =S_{0} \left( \mu  \right) ,\qquad A \left( 
\mu ,0 \right) = A_{0} \left( \mu  \right)\, .
\label{pab6}\end{equation} 
In our case the absence of the imaginary
unit from the exponential does not signal the sojourn of our spin in
forbidden terrain but simply accounts for the dissipative
character of the dynamics in consideration. Incidentally, due to the
presence of the large parameter $J$ even modest changes of $S_0$ are
reflected in wild fluctuations of $\rho(\mu,0)$; the ansatz therefore
does not limit our discussion to smooth  probability  distributions.

No loss of generality is incurred by assuming the function $S \left(
\mu ,\tau \right) $ independent of $J$ since the prefactor
$A \left( \mu ,\tau \right)$ may pick up all  dependence on $J$.
We represent the latter by an expansion in powers of
$\Delta =J^{-1}$
\begin{equation} 
A \left( \mu ,\tau \right) =A^{ \left( 0 \right) } \left( \mu ,\tau \right) 
+A^{ \left( 1 \right) } \left( \mu ,\tau \right) \Delta +A^{ \left( 
2 \right) }\Delta ^{2}+\ldots \,.
\label{pab61}\end{equation} 
The master equation (\ref{pab4}) then allows to determine
$S,A^{(0)}\ldots$ recursively. We shall need the equations for the
action and the zero-order prefactor,
\bea
{\partial S \over \partial \tau}+(1-\mu^2) \left[ 1-{\rm e}^{{\partial
 S \over \partial \mu}}\right]&=&0\,,
\label{pab10} \\
\left(\frac{\partial}{\partial\tau}-{\rm e}^{{\partial S
\over\partial\mu}}
(1-\mu^2)\frac{\partial}{\partial\mu}\right)\ln A^{(0)}
&=& {\rm e}^{{\partial S \over\partial\mu}}\left[ (1-\mu^2){1 \over 2}
{\partial 
 ^{2}S \over \partial \mu ^{2}} -\mu  \right] -\mu\, .
\label{pab11}
\eea
We shall neglect all higher-order corrections to the zero-order
prefactor.

\paragraph{Hamiltonian dynamics ---}
We may consider (\ref{pab10}) as the Hamilton-Jacobi equation for a
classical system with one degree of freedom and the Hamiltonian function
\begin{equation} 
H \left( \mu ,p \right) = (1-\mu^2) \left( 1-\E^{p} \right) .
\label{pab12}\end{equation}
The canonical equations of motion
$
\dot{\mu }= {\partial H \over \partial p}=-(1-\mu^2)\E^{p},\qquad
\dot{p}=-{\partial 
 H \over \partial \mu }=2\mu  \left( 1-\E^{p} \right) 
$ 
are easily integrated. In the resulting ``Hamiltonian'' trajectories,
\bea 
\tau ={1 \over 2a} \ln  { \left( a+\nu  \right)  \left( a-\mu  \right)
 \over  \left( a-\nu  \right)  \left( a+\mu  \right) }\,, \label{pab131}\\
 p=\ln  {a^{2}-\mu ^{2} \over 1-\mu ^{2}}\,,\label {pab131p}
\eea 
we denote by $\nu$ the initial coordinate; the name ``Hamiltonian'' is meant
to distinguish these solutions from the trajectories of the overdamped
pendulum (see below). The second integration constant, 
$a$, determines the ``energy'' $E=H(\mu,p)$ as
\begin{equation} 
a\equiv \sqrt{1-E} \,.
\label{pab14}
\end{equation} 

Rather remarkably, the Hamiltonian trajectory \rf{pab131} coincides
with the saddle-point equation incurred in I when examining the
asymptotics of the Laplace representation of the propagator. The
saddle-point parameter $a$  reappears in a new ``energetic'' role.
For later reference we note the nonnegative  ``speed''
\be
\dot{\mu}=-(a^2-\mu^2) \,.
\label{pab1369}
\ee 

A special  class of trajectories has zero initial momentum,
$p(\tau=0)=0$, therefore vanishing energy $E$ and $a=1$.  Their
Hamiltonian trajectories,
\begin{equation} 
\tau ={1 \over 2} \ln  { \left( 1+\nu  \right)  \left( 1-\mu  \right) 
 \over  \left( 1-\nu  \right)  \left( 1+\mu  \right) },\qquad p(t)= 0\,,
\label{pab1311}
\end{equation}  
are in fact just those of the classical overdamped pendulum
(\ref{pendulum}), disguised by $\mu=\cos\theta$. They involve the 
canonical momentum as conserved with the value zero.

The semiclassical quantum effects which our Hamiltonian
dynamics imparts to the spin through the WKB ansatz (\ref{pab5}) may
be seen in the existence of the Hamiltonian trajectories (\ref{pab131})
not included in the special class (\ref{pab1311}).

\paragraph{Solution of the Hamilton-Jacobi equation ---}
The familiar relation between canonical momentum and action,
\begin{equation} 
p={\partial S \left( \mu ,\tau \right)  \over \partial \mu }\,,
\label{pab121}\end{equation} 
implies
$ 
p_{0} \left( \nu  \right) ={\partial S_{0} \left( \nu  \right)  /
 \partial \nu }
$
at the initial moment $\tau=0$.
Since $ S_{0}$ is fixed by the initial density distribution this
latter equation uniquely associates an initial momentum with the initial
coordinate $\nu$.  One and only one Hamiltonian trajectory
$\mu(\tau;\nu,a)$ thus passes
at $\tau=0$ through the initial coordinate $\nu$, provided of course
that we consider the initial probabilities as imposed.
Conversely, we can find the initial coordinate  $\nu=\nu(\mu,\tau)$ from
which the current coordinate $\mu$ is reached at time $\tau$ along the
unique Hamiltonian trajectory.

The action $S(\mu,\tau)$ can now be obtained by integration along the
trajectory just discussed,
\begin{equation} 
S \left( \mu ,\tau \right) =\lt[S_{0} \left( \nu  \right) +\int^{\mu}_{\nu}p
  d\mu -E\tau\rt]_{\nu=\nu(\mu,\tau)}.
\label{pab18}\end{equation} 
We use the explicit form of the Hamiltonian trajectories
(\ref{pab131p}) to do the integral. For the sake of later convenience
we express the resulting action in terms of the auxiliary functions
\bea
 \sigma(a,\mu,\nu)&=& \;\;\;(\nu+a)\ln(\nu+a)-(\mu+a)\ln(\mu+a)\no\\
& & -(a-\nu)\ln(a-\nu)+(a-\mu)\ln(a-\mu)\label{sigma}\,,\\
\Phi(\mu,\nu,\tau)&=& \left[ \sigma(1,\mu,\nu)
-\sigma(a,\mu,\nu)+\tau(a^2-1)\right]_{a=a(\mu,\nu,\tau)}
\label{phidef}
\eea
as
\be
S(\mu,\tau)=\lt[S_{0} \left( \nu  \right)
+\Phi(\mu,\nu,\tau)\rt]_{\nu=\nu(\mu,\tau)}. 
\label{explicitaction}
\ee
In the definition (\ref{phidef}) of the function $\Phi$ the parameter
$a$ must, as indicated in the notation above, be read as a function of the
initial and final values of the coordinate since these are at
present considered as defining a Hamiltonian trajectory.
We may interprete the function $\Phi$ as the action accumulated along the
Hamiltonian trajectory in question. Its derivatives with respect to $\mu$ and
$\nu$ accordingly give the final and initial momenta,
\bea
\frac {\ptd \Phi}{\ptd \mu}=\ln \frac {a^2-\mu^2}{1-\mu^2}=p,\label{fomca}\\
\frac {\ptd \Phi}{\ptd \nu}=-\ln \frac
{a^2-\nu^2}{1-\nu^2}=-p_0\,.\label{famca}
\eea
The Hamiltonian trajectory $\mu(\tau;\nu,a)$ can be regarded as the solution
with respect to $\mu$ of the equation
\be
\frac {\ptd \Phi(\mu,\nu,\tau)}{\ptd \nu}=- \frac {\ptd S_0(\nu)}{\ptd \nu}.
\label{pab158}
\ee
 
\paragraph{WKB prefactor ---}
The expression \rf{pab11} for the prefactor can be simplified using the
notion of the full time derivative of a function $f(\mu,\tau)$ along the
Hamiltonian trajectory $\mu(\tau;\nu,a)$,
$
{df \left( \mu ,\tau \right)  \over d\tau} =\left. {\partial f \left( 
\mu\left(\tau,\nu\right),\tau\right)\over\partial\tau}\right|_{\nu
}=\left.{\partial f\over\partial\tau}\right|_{\mu }+\left.\dot{\mu}
 {\partial f \over \partial \mu }\right|_{\tau}\,,$
since the left hand side in (\ref{pab11}) is just the full time
derivative $dA /dt$ [see Eq.(\ref{pab121})].
We next introduce the Jacobian
\begin{equation} 
Y\left(\tau;\nu,a\right)={\partial\mu\left(\tau,\nu\right)\over\partial
\nu} ,\qquad Y \left( 0,\nu  \right) =1,
\label{pab21}\end{equation} 
and a new exponent $B(\mu ,\tau)$ to rewrite the prefactor as
\begin{equation} 
A ={\E^{B \left( \mu ,\tau \right) } \over \sqrt{Y}} \,.
\label{pab20}\end{equation} 
The full time derivative of $Y$  can be transformed to    
\bea
{dY \over d\tau}&=& {\partial ^{2}\mu  \over \partial \tau\partial \nu }
={\partial\over \partial \nu } \dot{\mu }={\partial  \over \partial \nu }
{\partial H(\mu,p) \over \partial p}={\partial \mu  \over \partial \nu }
\left({\partial^{2}H\over\partial\mu\partial p}+{\partial p\over\partial\mu
 } {\partial ^{2}H \over \partial p^{2}} \right) \no\\
&=&Y \left[  {\partial ^{2}H \over \partial \mu \partial p}+ {\partial
^{2}S\left(\mu ,\tau\right)\over\partial\mu^{2}}{\partial^{2}H\over\partial
p^{2}}\right]=Y\exp\left({\partial S\over\partial\mu}\right)
\left[2\mu -(1-\mu^2){\partial^{2}S\over\partial\mu^{2}}\right]\, .
\label{pab22}
\eea
So equipped we find the simple evolution equation ${dB \over
d\tau}=-\mu$ for the function $B(\mu,\tau)$ which can be integrated
along the trajectory to give
\begin{equation} 
B\left(\mu ,\tau\right)=-\int^{\tau}_{0}\mu d\tau +\ln A\left(
\nu ,0\right)=-{1\over 2}\ln {a^{2}-\mu^{2}\over a^{2}-\nu^{2}}+\ln A(\nu,0).
\label{pab25}\end{equation} 
We thus arrive at the asymptotic solution of the Cauchy problem for our
master equation with the  initial
condition \rf{pab6},
\begin{equation} 
\rho  \left( \mu ,\tau \right) = {1 \over \sqrt{{\partial \mu  \left( 
\tau,\nu  \right)  \over \partial \nu }}} \sqrt{{a^{2} -\nu ^{2} \over
a^{2}-\mu^{2}}}\E^{J\Phi\left(\mu
,\nu,\tau\right)}\rho\left(\nu,0\right)\,,
\label{pab27}\end{equation} 
where $\nu,a $ are meant as functions of $\mu $ and $\tau$ as
explained above.

\paragraph{Narrow versus broad initial distributions\label{disc} ---}
We have in effect constructed the solution \rf{pab27} by the method of
characteristics. At any rate, the  density  at a certain point is obtained
from the initial density by transport along the Hamiltonian
trajectory and acquires a factor consisting of an exponential and a
prefactor.

Let us show that the exponential factor is smaller than or at most equal to
unity.  According to \rf{fomca}  the extremum of $\Phi$ regarded  as a
function of the final coordinate $\mu$ with $\nu,\tau $ fixed occurs when
$a=1$, i.e. when $\mu$ moves like for the overdamped pendulum,
$\mu(\tau;\nu,a=1)\equiv\mu_{pend}(\tau;\nu)$ as defined by
\rf{pab1311} or (\ref{pendulum}). At the extremum we have $\Phi=0$ and
\be
\lt.\frac {\ptd^2 \Phi(\mu,\nu,\tau)}
{\ptd\mu^2 }
\rt|_{\mu=\mu_{pend}}
=-\lt.\frac{\Xi}
{(1-\mu^2)}
\rt|_{\mu=\mu_{pend}}
\label{secder}
\ee
where  $\Xi$ is a positive function defined as
 \be
\Xi(\mu,\nu,\tau)=\lt.\frac 1 {2a^2}\lt(\tau+\frac {\nu}{a^2-\nu^2}-
\frac {\mu}{a^2-\mu^2}\rt)\rt|_{a=a(\mu,\nu,\tau)}.
\label{defxi2}
\ee
But since $\Phi$ has a negative second derivative its
extremum is indeed a maximum, hence $\Phi\le 0$. We should mention that
by using \rf{secder} it is easy to calculate the integral of the density
\rf{pab27} over $\mu$ by the saddle-point method and to demonstrate that
our solution does not violate probability conservation.

Let us consider two extreme cases of the initial density distribution.
First suppose that the  initial density $\rho_0(\nu)$ is a smooth
function, with a gradient of order unity,
whence $S_{0}(\nu)$ vanishes. We would have $p_0=0, a=1$ which means
that the characteristic lines  are the overdamped-pendulum
trajectories (\ref{pab1311}). The time evolution (\ref{pab27})
then becomes the fully classical one,
\begin{equation} 
\rho\left(\mu ,\tau\right)=\left .{1-\nu^{2}\over 1-\mu^{2}}
\rho\left(\nu ,0 \right)\right|_{\nu=\nu_{class}(\mu,\tau)}\,.
\label{pab29}
\end{equation}
The speed of probability transport in this case is
obtained by putting $a=1$ in \rf{pab1369}; since that speed depends on
the coordinate the initial density distribution will change its form in
time, due to its finite spatial size. In particular, that change involves
broadening or sharpening depending on whether the distribution resides
mostly over positive or negative values of $\mu$, respectively. 

Now consider the opposite extreme of a narrow initial density, perhaps
one almost resembling a delta function. Then different parts of the
packet will have practically the same coordinate but highly different
momenta since
$
p_0\approx\frac 1 {J \rho_0}\frac {\ptd\rho_0}{\ptd \nu}.
$
The maximum has $p_0=0$ and thus moves along an overdamped-pendulum
trajectory. However the parts
on the left and right slope have, respectively, $p_0>0 $ and $p_0<0$.
They will be transported along the Hamiltonian trajectories with
$a>1$ and $a<1$, respectively. But according to \rf{pab1369} that means
that the density on the left slope travels in the direction of negative
$\mu$ faster than the one on the right slope. This will result in a
spreading of the initially narrow distribution. There is
an obvious superficial analogy with the decay of a wavepacket described
by the Schr\"{o}dinger equation. Of course, in our dissipative case the
exponential factor does not describe any dephasing but rather a
suppression
of probability propagation
along trajectories too strongly different from the fully classical
overdamped-pendulum ones.
This puts a brake on the spreading as soon as the packet widens;
quantitative estimates of the width will be given below in the discussion
of the properties of the propagator.  

As an example, in Fig. \ref{figevol} we demonstrate the  time dependence of
the probability distribution in the case when  the system was initially in a
pure coherent state of the angular momentum $\ket{\gamma}$ (for properties
of such states see e.g. \cite{Haake2,Arecchi}). The parameter $\gamma$
determines the direction of the mean spin vector $\av{\gamma|\bf J|\gamma}$
as $\gamma=\tan \frac{\theta_0}{2}\E^{i\phi_0}$. We took $j=200,
\gamma=0.4$. Three types of results are presented using: 
\begin{enumerate} 
\item{numerical integration of the master equation (''exact values``);}
\item{the WKB  solution \rf{pab27}; }
\item{the fully classical evolution formula \rf{pab29}.}
\end{enumerate}
	The  probability distributions given by the WKB formula coincide
	with the exact ones with accuracy in the range 0.4\%--  1.7\%
	(accuracy decreases at the later stages of the evolution).  The
	corresponding  plots  are indistinguishable. On the other hand, the
	fully classical formula  correctly places the probability peaks but
	grossly underestimates their broadening with time  leading to
	20\%--150\% error in the width and amplitude; this error does 
	not diminish as $j$ grows(!)

\paragraph{The semiclassical dissipative propagator\label{secprop} ---}
The dissipative propagator establishes a linear relation between the
initial and final density matrix elements. In the limit of large $J$ the
sum in this relation can be replaced by an integral; using the classical
variables $\mu,\nu$ as arguments it can be written (case $k=0$)
\begin{equation} 
\rho  \left( \mu ,\tau \right) =\int_{-1}^1 d\nu D\left( \mu ,\nu ,\tau
\right)  
 \rho  \left( \nu ,0\right)    
\label{pab30}\end{equation} 
with the function $D(\mu,\nu,\tau)$ related to the matrix $D_{mn}(\tau)$ as
$D \left(\mu ,\nu ,\tau \right)=J\left.D_{mn}(\tau)\right|_{m=J\mu,n=J\nu}$.

To obtain the propagator one has to solve the master equation with the
$\delta$-peak as the initial density distribution. Strictly speaking
such an initial condition does not fall into the class \rf{pab5} such
that our solution of the Cauchy problem \rf{pab27} is not directly
applicable. It is easy, however, to extract the propagator out of
\rf{pab27} in a slightly roundabout way. The semiclassical solution of
the dissipative problem in the form (\ref{pab5}) points to an analogy
between our master equation for the densities and a Schr\"{o}dinger
equation in imaginary time. In the spirit of that analogy we may
consider the function $D(\mu,\nu,\tau)$ in (\ref{pab30}) as the van Vleck
propagator \cite{Gutzwiller1} which must have the structure
\begin{equation} 
D\left( \mu ,\nu ,\tau \right) =R ( \mu ,\nu ,\tau ) \E^{J\Phi \left( 
\mu ,\nu ,\tau \right) }\,,
\label{pab31}\end{equation}
with $\Phi\left( \mu ,\nu ,\tau \right)$ the action accumulated along
the trajectory.

Our task is to establish the prefactor $R$. To do so let
us substitute the initial density  (\ref{pab5}) and the
semiclassical propagator in the form \rf{pab31} into (\ref{pab30}) and
perform the integration in the saddle-point approximation.
The maximum $\nu ^{*}=\nu ^{*}(\mu ,\tau)$ of the exponent
just defines the Hamiltonian trajectories in the form \rf{pab158}. The
saddle-point integration thus gives
\begin{equation} 
\rho  \left( \mu ,\tau \right) =R( \mu ,\nu ,\tau )
\sqrt{2\pi\over J}\left\{
-{\partial ^{2} \left[ \Phi \left( \mu ,\nu ,\tau \right) +S_{0}
\left( \nu  \right)  \right] \over \partial \nu ^{2}}
\right\}^{-1/2} \E^{J\Phi\left(\mu,\nu,\tau\right)}\rho\left(\nu,0
  \right) 
\label{pab33}
\end{equation} 
where $\nu^*(\mu,\tau)$ should be substituted for $\nu$.
Comparing with (\ref{pab27}) we find the  prefactor,
\begin{equation} 
R= \sqrt{\frac J {2\pi}}\left\{ - {{\partial  ^{2} \over \partial \nu ^{2}}
\left[ \Phi \left(  
\mu ,\nu ,\tau \right) +S_{0} \left( \nu  \right)  \right]   
 }\right\} ^{1/2} {\sqrt{{a^{2} -\nu ^{2} \over a^{2} -\mu ^{2}}} \over
 \sqrt{{\partial \mu  \left( \tau,\nu  \right)  \over \partial \nu
 }}}\,.
\label{pab34}\end{equation} 
A simpler form results once we realize from \rf{pab158} the Jacobian
to satisfy
\begin{equation}
{\partial \mu  \over \partial \nu }= - {{\partial  ^{2} \over \partial
 \nu ^{2}}  \left[ \Phi \left( \mu ,\nu ,\tau \right) +S_{0} \left( \nu
  \right)  \right]  \over {\partial ^{2}\Phi \left( \mu ,\nu ,\tau \right) 
 \over \partial \nu \partial \mu }}\,.
\label{pab35}\end{equation} 
For the propagator we thus find
\begin{equation} 
D\left( \mu ,\nu ,\tau \right) = \left[ {J \over 2\pi } {\partial
 ^{2}\Phi  \left( \mu ,\nu ,\tau \right)  \over \partial \nu \partial
 \mu } \right] ^{1/2}\sqrt{{a^{2} -\nu ^{2} \over a^{2} -\mu ^{2}}}
 \E^{J\Phi \left( \mu ,\nu ,\tau \right) }\,.
\label{pab36}\end{equation} 
Compared with the van Vleck propagator for the Schr\"{o}dinger equation
\cite{Gutzwiller1} the preexponential in this expression contains an
additional square root factor, the origin of which can be traced to the
difference in the normalizing condition for wave functions and density
matrices. It is system specific in as much as $ a$ is a solution of 
(\ref{pab131p}). Note, however, that on the classical trajectory ($a=1$) this
factor is just the square root of the classical jacobian
$d\mu^{-1}(\mu)/d\mu$, with $\nu=\mu^{-1}(\mu)$ the inverted classical
trajectory (\ref{pab1311}). One easily verifies that both square roots in
(\ref{pab36}) give rise 
to the same factor and combine to the jacobian to the power one,
as it is neccessary to guarantee probability conservation.  

By expressing the mixed derivative of the action $\Phi$ in the preexponent
in terms of the function $\Xi(\mu,\nu,\tau)$  introduced in \rf{defxi2},
\be\label{pab66}
\frac {\ptd ^2\Phi(\mu,\nu,\tau)}{\ptd \mu \ptd \nu}=  
\lt[(a^2-\mu^2)(a^2-\nu^2)\;\Xi(\mu,\nu,\tau)\;\rt]^{-1} \,
\ee 
we arrive at the final form of our semiclassical density propagator matrix
$D_{mn}$,
\be
D_{mn}(\tau)=\lt .\frac 1 {(a^2-\mu^2)\sqrt{2\pi J\Xi\;}}\E^{J\Phi(\mu,\nu,
\tau)}\rt|_{\mu=\frac m J;\;\nu=\frac n J}\,,
\label{asymd}
\ee 
in which $a$ must be read as the function $a=a(\mu,\nu,\tau)$ since a
Hamiltonian tractory is already uniquely determined by specifying the
initial and final coordinates. While in general that function can be found
only numerically from the Hamiltonian trajectories \rf{pab131}, certain
limits do yield approximate analytical  expressions (see below).
Note that  $a$ is always larger than the larger of $|\mu|,|\nu|$.

The behaviour of the propagator solution is largely determined by the action
$\Phi$. Two situations deserve special mention.   
First suppose that  the initial
quantum number $n$  is not close to $j$. Then we encounter a narrow packet
centered around the ``fully classical'' final point
$\mu_{pend}(\tau;\nu)$ where the maximum of $\Phi$ is located.  Close to
that maximum the propagator can be represented by the Gaussian
\be
D_{mn}(\tau)\approx  
\frac 1 {d \sqrt{2\pi}}
\exp\lt[
-\frac{\lt(\mu-\mu_{pend}\rt)^2}
{2d^2}
\rt].
\label{gauss}
\ee
whose width  $d(\nu,\tau)$ is determined by the second derivative of the
action \rf{secder}. If expressed in terms of the classical coordinate $\mu$
the width $
d=\lt[-J\Phi_{\mu\mu}(\mu_{pend},\nu,\tau)\rt]^{-1/2}
$
goes to zero in the classical limit like $J^{-1/2}$,
such that for many purposes the propagator \rf{gauss} may even be identified
with the $\delta$-function. 
  
A radically different situation is met with when the initial quantum number
$n$ is close or equal to $j$ so that the initial coordinate $\nu$ tends to 1
when $J\to\infty$. The action accumulated  along any Hamiltonian
trajectory is then close to zero like $1-\nu$  and so is
its  second derivative. Indeed, one easily checks what we already saw in
I,
\bea
a&\approx& 1-\lt(1-\E^{-2\tau'}\rt)\delta_{\nu} \quad{\rm with} \\
\delta_{\nu}&=&1-\nu=(j-n)/J \,
\eea
where $\tau'=\tau-\tau_{class}(\mu,\nu)$ denotes a quantum time shift,
i.e. the difference between the travel times along the Hamiltonian
trajectories (\ref{pab131}) and the overdamped-pendulum ones (\ref{pab1311}).
The action then comes out as
\be
\Phi\approx \delta_{\nu}\lt(1-2\tau'-\E^{-2\tau'}\rt).
\ee 
Obviously, the exponent $J\Phi$ tends to a finite function of the
relative time $\tau'$ when $J$ tends to infinity as $j-n$ is kept fixed.
The exponential factor is then only
slightly less than unity along a Hamiltonian trajectory. Thus the
width of the distribution does not tend to zero but rather stays
finite as $J\to\infty$.

\paragraph{Comparison of the semiclassical propagators ---}
The semiclassical propagator obtained in I by the saddle-point method,
\be
D_{mn}(\tau)=\frac 1 {(1-\mu^2)\sqrt{2\pi
J\Xi\;}}\E^{J\Phi(\mu,\nu,\tau)}\,.
\label{asymd2}
\ee
intriguingly differs from our present \rf{asymd} in two respects. First,
instead of $a^2-\mu^2$ in the prefactor, \rf{asymd2} contains $1-\mu^2$.
Second, in \rf{asymd2} and in the whole of I we 
connected the classical coordinate  $\mu$ to the quantum number $m$ by
$
\mu=\frac {m-1} J
$,
which differs from our present definition   $\mu=m/J$  by the small shift
$\Delta=J^{-1}$.

Let us show that these two changes in fact cancel each other to leading
order in $J^{-1}$.  Consider the form \rf{asymd} of our present
propagator and substitute $\mu=\mu'+\Delta$ with $\mu'$ the
shifted argument $(m-1)/J$. If $m$ is not very close to $n$ the
denominator in the prefactor is not small and one can neglect its
change brought about by the replacement $\mu\rightarrow\mu'$. As regards
the exponent we must exercize greater care: The large factor $J$
obliges us to keep two terms in the expansion
\be
J\Phi(\mu,\nu,\tau)=J\Phi(\mu'+\Delta,\nu,\tau)=J\Phi(\mu',\nu,\tau)
+\ln \frac{a^2-\mu^2}{1-\mu^2}+{\cal O}(\Delta)\,.
\ee
The logarithm arising here obviously modifies the prefactor just so as
to bring about the propagator (\ref{asymd2}) of the previous
paper. Therefore, for $J\to\infty$ and $m$ not close to $n$, the two forms
(\ref{asymd2}) and (\ref{asymd}) of the propagator are equivalent.

If $m$ approaches $n$ the equivalence of the formulae \rf{asymd} and
\rf{asymd2} does not hold any more: the accuracy of \rf{asymd2} is
higher. In the extreme cases when $n-m$ or $j-n$ or $j\pm m$ are of
order unity or zero \footnote{These cases are similar to the ground
state of a quantum system and cannot  be adequately described by the
WKB method.} the uniform approximation for the propagator should be used
instead of either \rf{asymd} or \rf{asymd2} [see I].

\section{Conclusion}
The role of the WKB approximation in quantum mechanics as a bridge
to classical mechanics is common
knowledge. What is not fully appreciated, though, is its
potential usefulness for dissipative problems described by master
equations. The problem of spin damping in superradiance may serve as a
good example.

All results of the present paper followed from the WKB ansatz \rf{pab5}.
It led us to a Hamilton-Jacobi equation of a classical mechanical system
with one degree of freedom. The two canonical variables satisfy two
 Hamilton equations which require two initial or boundary data
(like the initial coordinate and the initial momentum or the initial and
final coordinate) to define a trajectory. This contrasts with the
trajectories of the overdamped pendulum model described by a single
differential equation of first order which are uniquely defined by
the initial coordinate only. The extended family of Hamiltonian
trajectories suggests an intuitive explanation of the quantum broadening of
an initially sharp distributions alien to the
traditional classical model: Although all relevant classical
trajectrories have the same initial coordinate, the initial momenta
are different, and so are the final coordinate.

There is of course  a basic difference between the
spreading of wave functions  and the
dissipative propagation of probabilities. In the case of unitary evolution
the probability amplitudes to arrive from a fixed initial point to different
final points differ by the phase  accumulated along the respective classical
trajectory (that is, if we neglect the prefactor). Consequently, propagation
along all classical trajectories starting from the same initial point takes
place with a comparable probability. In our dissipative problem the exponent
of the WKB solution is either zero or negative. This means that there is no
equality among the Hamiltonian trajectories starting at the same initial
point: 
 the one which corresponds to vanishing action (the fully
classical overdamped-pendulum trajectory) is privileged to contribute
most. Probability propagation along trajectories
significantly removed from that privileged one is effectively
suppressed. As a result the final width of an initially sharp
distribution remains small, unlike the eventually infinite dispersion of the
wavepacket of a free particle. The exceptional character of the
superradiant evolution starting from the highest energy levels is also
easily understood. In that case the classical action is close to zero on
all trajectories. Therefore the exponential factor does not limit any
more the broadening of the distribution with time.

The necessity to extend the dynamics of the
overdamped-pendulum model was recognized long ago, see the reviews
\cite{Haroche,Benedict}. In \cite
{Haake1} a family of trajectories described by two first order
differential equations was considered (they were introduced as the
characteristics in the propagation of quasiprobability
distributions); the two variables there employed did not form
a canonical pair, though; one of them was related to our
present coordinate $\mu$ and the other to the transverse
components of the spin.

Our present WKB approach has the merit of being easily extended to other
problems in dissipative quantum mechanics. Apart from providing an
intuitive qualitative picture involving a Hamiltonian equivalent it also
provides a convenient analytic approximation for the dissipative
propagator. In particular we used it to calculate the width of the final
distribution in the case when the initial direction of the Bloch vector
was not close to the $z$ axis. This result is of little significance for
the theory of superradiance itself. However, the master equation
\rf{master} becomes increasingly important as a model in investigations of
quantum chaos in dissipative systems \cite{Grobe,bible}; disregard of
the final width of the propagator solutions would lead to erroneous
results there. We shall extend the present work to dynamics with chaos
in a subsequent paper.

\section{Appendix. Propagation of coherences}

No separate investigation of the coherence propagator $D_{mn}^k(\tau)$ is
necessary  because of  the identity proven in I ,
\be
D_{mn}^k(\tau)=D_{mn}(\tau)\frac{\sqrt{Q_{m-k,n-k}Q_{m+k,n+k}}}{Q_{mn}}
\E^{ k^2 \tau/J}\label{identity}
\ee with
\be
Q_{mn}=\prod_{l=m+1}^n g_l=\frac{(j+n)!(j-m)!}{(j+m)!(j-n)!}.
\label{defQ}
\ee
It is instructive, however, to consider the  changes in our Hamilton-Jacobi
formalism necessitated by nonzero $k$.
The new quantum number $k$ whose range goes to infinity when
$j\to\infty$ must be accompanied with a macroscopic variable
\be
\eta=\frac k J\,.
\ee

The master equation \rf{master} written with $\mu,\nu$ as arguments
now reads
\bea
J\frac {\ptd \rho(\eta; \mu,\tau)}{\ptd \tau}
&=&J^2\sqrt{\lt[1-(\mu+\eta)^2-(\mu+\eta)\Delta\rt]\lt[1-(\mu-\eta)^2-
(\mu-\eta)\Delta\rt]}\rho(\eta;\mu+\Delta,\tau)\no\\
&&-J^2\lt(1-\mu^2-\eta^2+\mu\Delta\rt)\rho(\eta; \mu,\tau)+{\cal O}(\Delta^2).
\eea
A Hamilton-Jacobi ansatz
\be
\rho(\eta;\mu,\tau)=A(\eta;\mu,\tau)\E^{JS(\eta;\mu,\tau)}
\ee 
entails a chain of differential equations for the ``action'' $S$
and the terms in the expansion of the amplitude $A$ in powers of $\Delta$.
We shall examine only  the Hamilton-Jacobi equation
 \be
\frac{\ptd S}{\ptd \tau}+G(\mu)-F(\mu)\exp\lt(\frac{\ptd S}{\ptd \mu}\rt)=0
\ee
where $F$ and $G$ denote the auxiliary functions
\be
F(\mu)=\sqrt{[1-(\mu+\eta)^2][1-(\mu-\eta)^2]},\quad G(\mu)= 1-\mu^2-\eta^2.
\ee

The previous Hamiltonian becomes extended to
\be
H(\mu,p)=G(\mu)-F(\mu)\E^p.
\ee
Once more denoting the conserved value of $H$ by $E$ and introducing the
constant $a$ by the relation
\be
a=\sqrt{1-E-\eta^2}
\label{newadef}
\ee
we obtain the canonical equation for the coordinate,
\be
\dot{\mu}=-F\E^p=a^2-\mu^2\,,
\ee
It coincides with \rf{pab1369}, and its integration leads
to exactly the same trajectories \rf{pab131} as for the densities.
The characteristic lines for the coherences propagation are the same as
the ones for probability propagation.

There is one cardinal difference. For $k=0$ we could single out the
special trajectories with zero initial momenta. Only these were
important in the case of smooth initial densities, according to the
relation $p\approx\ptd \ln\rho/J \ptd \mu$. Since the initially vanishing
momentum remained  zero at $\tau\ne 0$, an initially smooth density
remained  smooth as long as $\tau$ was not too large (before the system
reached its lowest energy level).
Smooth density distributions therefore form a closed class, and the
overdamped-pendulum trajectories are their routes of propagation.

In the case of coherences it is also possible to select the trajectories
corresponding to smooth initial distributions or zero inital momenta:
the parameter $a$ should be chosen then according to
\be
a^2=1-\eta^2-G(\nu)+F(\nu)\,,
\ee
but now these trajectories do not have the physical significance of
the fully classical trajectories of the overdamped pendulum. The reason
simply is that an initially vanishing momentum $p$ will no longer be
zero when $\tau\ne 0$. Indeed, unless $\eta=0$ and $a=1$, the momentum
\be
p=\ln \frac {G(\mu)-E}{F(\mu)} =\ln \frac {a^2-\mu^2}{\sqrt{[1-
(\mu+\eta)^2][1-(\mu-\eta)^2]}},
\ee
 with $\mu=\mu(\tau)$ 
can never be a constant. Remembering the momentum-density connection we
conclude that an initially smooth distribution of coherences inevitably
ceases to be
smooth in the course of time.  Thus there is no special class of
characteristics responsible for transporting smooth distributions of
coherences and leaving them smooth, hence no elementary relation
like \rf{pab29}.

Our considerations throw light on the important question whether
it is admissible to replace the master equation by a
first-order differential equation using approximations like
\be
\rho_{m+1}\approx\rho_m+\frac {\ptd \rho_m}{\ptd m}
\label{repla}
\ee
Such a replacement is justified if the respective elements of the density
matrix are and remain smooth functions of $m$. It follows:
\begin{itemize}
\item{In the case of density propagation the above replacement \rf{repla}
is justified provided the initial  density distribution is smooth.}
\item{For coherences the replacement \rf{repla} is always illegal.}
\end{itemize}

\begin{figure}
\protect\caption{\label{figevol}}
Snapshots of the probability distribution $\rho(\mu,\tau)$ at various times,
for the initially pure coherent state 
($\gamma=0.4, j=200$). The WKB [Eq.(\ref{pab27})] and exact results shown by
filled contours coincide in the scale of the plot. Dashed contours
correspond to 
 the  classical evolution formula \rf{pab29} based on the dynamics of the
overdamped top. It  grossly underestimates the width and overestimates the
height of the peaks.  
\end{figure}

\end{document}